%% file: 1614-nature-letter.tex
\documentclass[12pt]{article}

\usepackage{graphicx}
\usepackage{aas_macros}
\usepackage{amssymb,amsmath}
\newcommand{\msun}{M$_\odot$}
\newcommand{\first}{1\textsuperscript{st}}
\newcommand{\second}{2\textsuperscript{nd}}
\newcommand\arcmin{\mbox{$^\prime$}}%
\newcommand{\farcs}{\mbox{\ensuremath{.\!\!^{\prime\prime}}}}%
\newcommand\fs{\mbox{$.\!\!^{\mathrm s}$}}%

\long\def\symbolfootnote[#1]#2{\begingroup%
\def\thefootnote{\fnsymbol{footnote}}\footnote[#1]{#2}\endgroup} 

\title{Shapiro delay measurement of a 2 solar mass neutron star}

\author{P.~Demorest$^{1}$, T.~Pennucci$^{2}$, S.~Ransom$^{1}$,
M.~Roberts$^{3}$ \& J.~W.~T.~Hessels$^{4,5}$}

\begin{document}

\maketitle

\begin{enumerate}
 \item National Radio Astronomy Observatory, 520 Edgemont Road, Charlottesville, VA 22093 USA
 \item Astronomy Department, University of Virginia,
 Charlottesville, VA 22094-4325 USA
 \item Eureka Scientific, Inc., Oakland, CA 94602, USA
 \item Netherlands Institute for Radio Astronomy (ASTRON), 
   Postbus 2, 7990 AA Dwingeloo, The Netherlands
 \item Astronomical Institute ``Anton Pannekoek,'' University 
   of Amsterdam, 1098 SJ Amsterdam, The Netherlands
\end{enumerate}

{\bf
  Neutron stars are composed of the densest form of matter known to
  exist in our universe, and thus provide a unique laboratory for
  exploring the properties of cold matter at supranuclear density.
  Measurements of the masses or radii of these objects can strongly
  constrain the neutron-star matter equation of state, and consequently
  the interior composition of neutron stars\cite{lp04,lp07}.  Neutron
  stars that are visible as millisecond radio pulsars are especially
  useful in this respect, as timing observations of the radio pulses
  provide an extremely precise probe of both the pulsar's motion and the
  surrounding space-time metric.  In particular, for a pulsar in a
  binary system, detection of the general relativistic Shapiro delay
  allows us to infer the masses of both the neutron star and its binary
  companion to high precision\cite{jhb+05,vbs+08}.  Here we present
  radio timing observations of the binary millisecond pulsar
  PSR~J1614$-$2230, which show a strong Shapiro delay signature.  The
  implied pulsar mass of 1.97$\pm$0.04\,M$_\odot$ is by far the highest
  yet measured with such certainty, and effectively rules out the
  presence of hyperons, bosons, or free quarks at densities comparable
  to the nuclear saturation density.  
}

In the accepted ``lighthouse model'' description of radio pulsars, a
rapidly spinning neutron star (NS) with a strong magnetic field
($10^8$-$10^{15}$\,G) emits a beam of radiation that is typically
misaligned with the spin axis.  A broadband, polarized pulse of radio
emission is observed once per rotation if this beam crosses the
Earth-pulsar line of sight.  The extraordinary rotational stability of
pulsars permits the precise measurement of a number of systematic
effects that alter the arrival times of the radio pulses at Earth, a
procedure referred to as pulsar timing.  In the case of binary
millisecond pulsars, which are the most stable pulsars with orbital
companions, even typically subtle effects such as the general
relativistic Shapiro delay can be revealed by timing.  The Shapiro delay
is an increase in light travel time through the curved space-time near a
massive body.  In binary pulsar systems that have highly inclined
(nearly edge-on) orbits, excess delay in the pulse arrival times can be
observed when the pulsar is situated nearly behind the companion during
orbital conjunction.  As described by general relativity, the two
physical parameters that characterize the Shapiro delay are the
companion mass and inclination angle.  In combination with the observed
Keplerian mass function, the Shapiro delay offers one of the most
precise methods to directly infer the mass of the NS.  In turn, any
precise NS mass measurement limits the equations of state (EOS)
available to describe matter at supranuclear densities.  The discovery
of a NS with mass significantly higher than the typical value of
$\sim$1.4\,\msun~would have a major impact on the allowed NS EOS as well
as additional implications for a wide range of astrophysical
phenomena\cite{opr+10}.

PSR~J1614$-$2230 was originally discovered in a radio survey of
unidentified EGRET gamma-ray sources using the Parkes radio
telescope\cite{hrr+05}.  The spin period $P$ is 3.15\,ms, and initial
timing with Parkes showed the pulsar to be in a binary system with an
8.7-day orbital period and a companion of mass $M_2 \gtrsim
0.4$\,M$_\odot$.  The system was noted as having a higher companion mass
than is typical for fully-recycled ($P\lesssim10$\,ms) pulsars, which
predominantly have helium white dwarf (WD) companions with masses of
$\sim$0.1$-$0.2\,\msun.  Furthermore, the orbital period is shorter than
expected given the massive companion\cite{rpj+95}.  These facts hinted
at a possible non-standard evolutionary history for the binary system
(see below).  

Following the inital discovery observations, J1614$-$2230 was observed
regularly as a test source for several observing projects at the NRAO
Green Bank Telescope (GBT)\cite{rrh+10}\symbolfootnote[2]{The National
Radio Astronomy Observatory is a facility of the U.S. National Science
Foundation (NSF), operated under cooperative agreement by Associated
Universities, Inc.}.  These data provide a continuous long-term timing
record from mid-2002 to the present, and they show a marginally
significant Shapiro delay signal.  However, the data quality and
arbitrary scheduling with respect to the binary's orbital phase made it
impossible to obtain meaningful mass and inclination measurements from
those data alone.  In March 2010, we performed a dense set of
observations of J1614$-$2230, timed to follow the system through one
complete orbit with special attention paid to the orbital conjunction,
where the Shapiro delay signal is strongest.  These data were taken with
the newly built Green Bank Ultimate Pulsar Processing
Instrument\cite{guppi}.  GUPPI coherently removes interstellar
dispersive smearing from the pulsar signal, and integrates the data
modulo the current apparent pulse period, producing a set of average
pulse profiles, or flux-versus-rotational-phase lightcurves.  We
observed an 800\,MHz wide band centered at a radio frequency of
1.5\,GHz.  The raw profiles were then polarization- and flux-calibrated
and averaged into 100-MHz, 7.5-minute intervals using the PSRCHIVE
software package\cite{psrchive}.  From these, pulse times of arrival
were determined using standard procedures, with a typical uncertainty of
$\sim$1\,$\mu$s.

The measured arrival times are used to determine key physical parameters
about the neutron star and its binary system by fitting them to a
comprehensive timing model which accounts for every rotation of the
neutron star over the time spanned by the fit.  The model predicts at
what times pulses should arrive at Earth, taking into account pulsar
rotation and spin-down, astrometric terms (sky position, proper motion),
binary orbital parameters, interstellar dispersion, and general
relativistic effects such as the Shapiro delay (see
Table~\ref{tab:pars}).  The observed arrival times are compared with the
model prediction, and best-fit parameters are obtained via $\chi^2$
minimization, using the TEMPO2 software
package\cite{tempo2}\symbolfootnote[3]{We also obtained consistent
results using the original TEMPO package.}.  The post-fit residuals,
i.e. the difference between the observed and model-predicted pulse
arrival times, effectively measure how well the timing model describes
the data, and are shown in Figure~\ref{fig:resid}.  We included both the
older long-term datasets and our new GUPPI data in a single fit.  The
long-term data determine model parameters with characteristic timescales
longer than a few weeks (e.g. spin-down rate and astrometry), while the
new data best constrain parameters on timescales of the orbital period
or less.  Both are necessary to reduce covariance between particular
model parameters.  

In addition to the physical timing model parameters listed in
Table~\ref{tab:pars}, the fit included arbitrary time offsets
(``jumps'') between different observing systems, and allowance for a
time-variable interstellar dispersion measure (DM), a quantity
proportional to the total free electron column density along the line of
sight.  A single DM value was fit for the long-term data set.  The
newer, more precise GUPPI data required a separate DM be fit for each
day.  DM is expected to vary on 1-day and longer timescales due to the
relative motions of the pulsar and Earth through the interstellar
medium\cite{rdb+06}.  We also investigated the possibility of DM
variation on minute to hour timescales which could indicate the presence
of excess gas in the binary system itself.  If present, especially near
orbital conjunction, this effect could bias the Shapiro delay
measurement.  However, we found no evidence of this in our data -- we
limit any such orbital DM variation to a level of $\lesssim
2\times10^{-4}$\,pc\,cm$^{-3}$, corresponding to delays of
$\lesssim$0.4\,$\mu$s at a frequency of 1.5\,GHz.  Furthermore, we
observed orbital conjunctions at three different epochs over a span of 2
months, and all show consistent timing results without needing to model
fast DM variation.

\begin{figure}
\includegraphics[width=\textwidth]{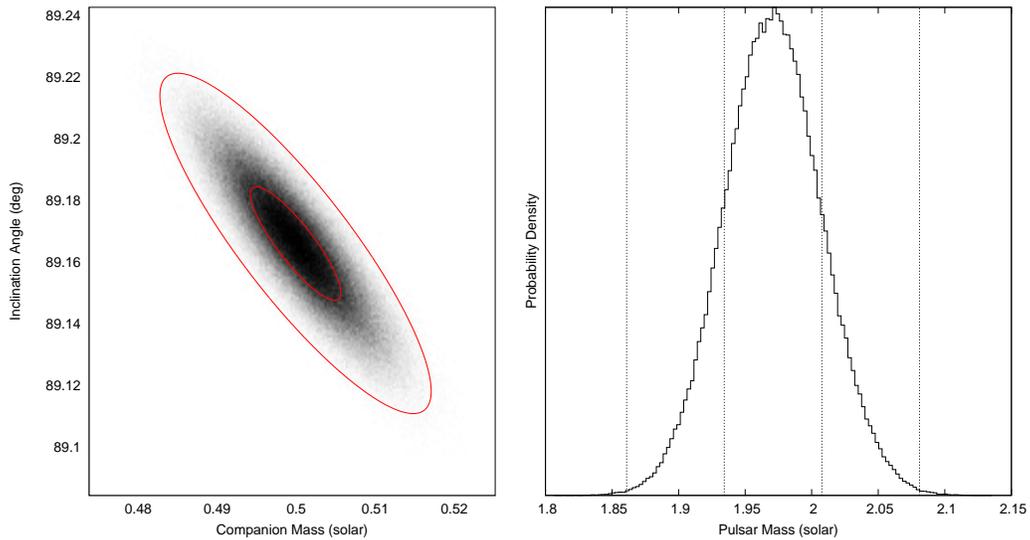}
  \caption{\label{fig:mcmc} Results of the Markov chain Monte Carlo
    (MCMC) error analysis.  The left panel grayscale shows the 2-D
    posterior probability density function (PDF) in the $M_2$-$i$
    (companion mass versus orbital inclination) plane computed from a
    histogram of MCMC trial values.  The ellipses show 1- and
    3-$\sigma$ contours based on a Gaussian approximation to the MCMC
    results.  The right panel shows the PDF for pulsar mass derived
    from the MCMC trials.  In both cases the results are very well
    described by normal distributions due to the extremely high
    signal-to-noise ratio of our Shapiro delay detection.}
\end{figure}

As shown in Figure~\ref{fig:resid}, the Shapiro delay was detected in
our data with extremely high significance, and must be included to
correctly model the arrival times of the radio pulses.  However,
estimating parameter values and uncertainties can be tricky due to the
high covariance between many orbital timing model terms, especially
the Shapiro-derived companion mass ($M_2$) and inclination angle
($i$)\cite{dd86}.  In order to obtain robust error estimates, we used
a Markov chain Monte Carlo (MCMC) approach to explore the post-fit
$\chi^2$ space and derive posterior probability distributions for the
model parameters (see Figure~\ref{fig:mcmc}).  In contrast with the
common method of mapping $\chi^2$ projections in a reduced-dimensional
parameter space\cite{vbb+02,jhb+05}, MCMC is able to efficiently
explore all fit dimensions simultaneously.  Our final results for all
timing model parameters, with MCMC error estimates, are given in
Table~\ref{tab:pars}.  Due to the high significance of the detection,
our MCMC procedure and the standard $\chi^2$ fit produce similar
uncertainties.

From the detected Shapiro delay we measure a companion mass of
$0.500\pm0.006$\,\msun, which implies that the companion is a
helium-carbon-oxygen white dwarf, as helium WD stars have masses of
$\lesssim0.45$\,\msun\cite{kw90}.  Considering the 8.7-day orbital
period, a likely formation mechanism for systems similar to
J1614$-$2230 has been proposed\cite{pr00,prp02} where the original
companion star is $\sim$3\,\msun.  J1614$-$2230 is peculiar, however,
in being fully recycled unlike the mildly-recycled pulsars
($P\sim10-30$\,ms) found in similar binary systems.

The Shapiro delay also shows the binary system to be amazingly edge-on,
with an inclination of $89.17^\circ \pm 0.02^\circ$.  This is the most
inclined pulsar binary system currently known. The amplitude and
sharpness of the Shapiro delay increase rapidly with increasing binary
inclination and the overall scaling of the signal is linearly
proportional to the mass of the companion star.  Thus the unique
combination of the high orbital inclination and massive WD companion in
J1614$-$2230 cause a Shapiro delay amplitude orders of magnitude larger
than for most other millisecond pulsars with 0.1$-$0.2\,\msun~companions
and inclinations $\lesssim$75$^\circ$.  In addition, the excellent
timing precision achievable from the pulsar with the GBT and GUPPI
provide a very high signal-to-noise ratio measurement of both Shapiro
delay parameters within a single orbit.

The standard Keplerian orbital parameters, combined with known companion
mass and orbital inclination, fully describe the dynamics of a ``clean''
binary system\symbolfootnote[4]{A system comprised of two stable compact
objects.} under general relativity and therefore also determine the
pulsar's mass.  We measure a NS mass of $1.97 \pm 0.04$\,\msun, by far
the highest precisely measured NS mass determined to date.  In constrast
with X-ray based NS mass/radius measurements\cite{ozel06}, the Shapiro
delay provides no information about the NS radius.  However, unlike the
X-ray methods, our result is nearly model-independent, as it depends
only on general relativity being an adequate description of gravity.  In
addition, unlike statistical pulsar mass determinations based on
measurement of the advance of periastron\cite{rhs+05,frb+08,fwb+08},
pure Shapiro delay mass measurements involve no assumptions about
classical contributions to periastron advance or the distribution of
orbital inclinations.  

\begin{figure}
\includegraphics[width=\textwidth]{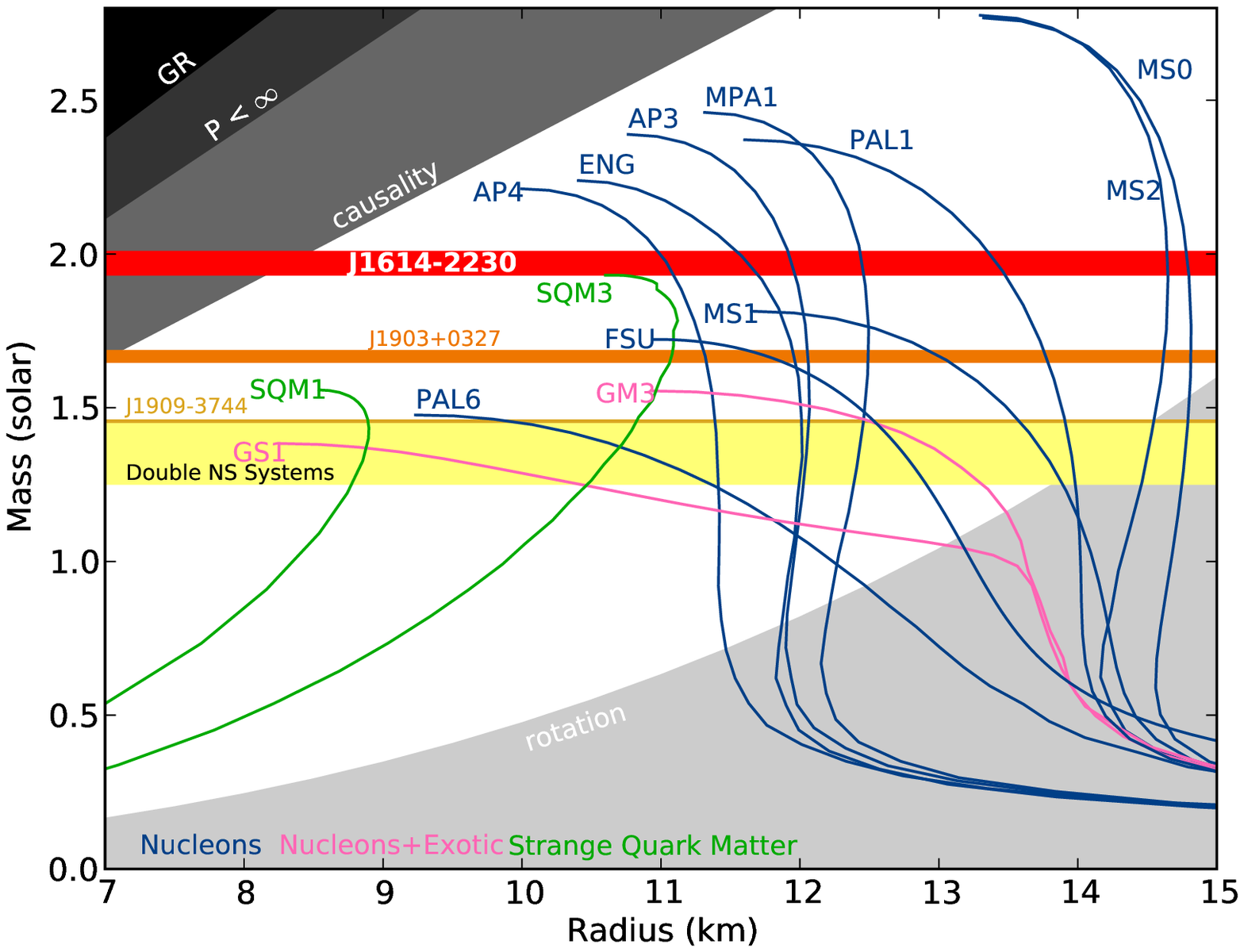}
  \caption{\label{fig:eos} Neutron star (NS) mass-radius diagram.  The
  plot shows non-rotating mass versus physical radius for several
  typical NS equations of state (EOS)\cite{lp01}.  The horizontal bands
  show the observational constraint from our J1614$-$2230 mass
  measurement of 1.97$\pm$0.04~\msun, similar measurements for two
  other millsecond pulsars\cite{jhb+05,1903}, and the range of observed
  masses for double NS binaries\cite{lp07}.  Any EOS line that does
  not intersect the J1614$-$2230 band is ruled out by this measurement.  In
  particular, most EOS curves involving exotic matter, such as kaon
  condensates or hyperons, tend to predict maximum NS masses well below
  2.0~\msun, and are therefore ruled out.}
\end{figure}

The mass measurement alone of a 1.97\,\msun\ NS significantly constrains
the NS EOS, as shown in Figure~\ref{fig:eos}.  Any
proposed EOS whose mass-radius track does not intersect the J1614$-$2230
mass line is ruled out by this measurement.  The EOSs which produce the
lowest maximum masses tend to be those which predict significant
softening past a certain central density.  This is a common feature of
models that include the appearance of ``exotic'' hadronic matter such as
hyperons or kaon condensates at densities of a few times the nuclear
saturation density ($n_s$), for example GS1 and GM3 in
Figure~\ref{fig:eos}.  All such EOSs are ruled out by our results.  Our
mass measurement does not completely rule out condensed quark matter as
a component of the NS interior\cite{krv10}, but it does constrain quark
matter model parameters, and excludes some strange quark matter
models\cite{opr+10}.  For the range of allowed EOS lines presented in
Figure~\ref{fig:eos}, typical values for the physical parameters of
J1614$-$2230 are a central baryon density between 2$-$5\,$n_s$ and a
radius between 11$-$15\,km, only 2$-$3 times the Schwarzschild radius
for a 1.97\,\msun~star.  Based on EOS-independent analytic solutions of
Einstein's equations, our mass measurement also sets an upper limit on
the maximum possible mass density of cold matter\cite{lp05}, of
$\lesssim4\times10^{15}$~g~cm$^{-3}\sim$10~$n_s$.

The proposed formation mechanism for J1614$-$2230-like
systems\cite{pr00,prp02} predicts that the neutron star would only
accrete a few hundredths of a solar mass of material, much less than the
$\sim$0.6\,\msun\ needed to bring a neutron star born at 1.4\,\msun\ up
to the observed pulsar mass.  This implies either that the transfer of
large amounts of mass onto the neutron star from the secondary is
possible or alternatively that some neutron stars are born massive
($\sim$1.9\,\msun).  As the Shapiro delay has not been detected in most
millisecond pulsar systems, our mass measurement for J1614$-$2230
suggests that many of these other systems may also harbor NSs with
masses well above 1.4\,\msun.

\pagebreak
\begin{center}
\includegraphics[width=\textwidth]{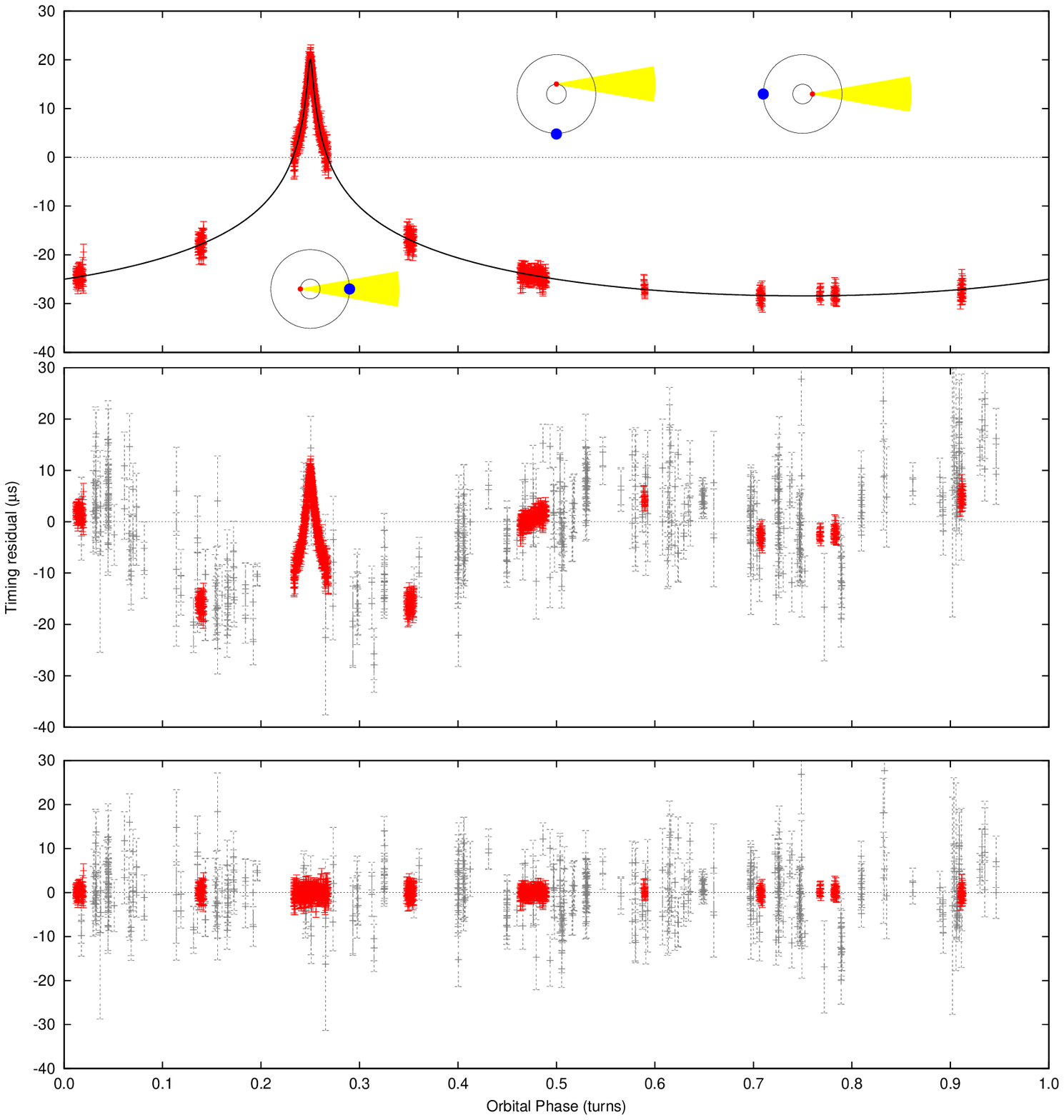}
\end{center}
\pagebreak

\begin{figure} 
  \caption{\label{fig:resid} Shapiro delay measurement for
    PSR~J1614$-$2230.  Each panel shows timing residual -- excess
    delay not accounted for by the timing model -- as a function of
    the pulsar's orbital phase.  The top panel shows the full
    magnitude of the Shapiro delay if it is not included in the timing
    model but with all other model parameters fixed at their best-fit
    values.  The solid line shows the Shapiro delay functional form,
    and the red points are timing measurements from our GBT/GUPPI data
    set.  The diagrams inset in this panel show a top-down schematic
    view of the binary system at orbital phases 0.25, 0.5, and 0.75.
    The neutron star is shown in red, the white dwarf companion in
    blue, and the emitted radio beam (yellow) points towards the
    Earth.  At orbital phase 0.25, the Earth--pulsar line of sight
    passes nearest to the companion ($\sim$240,000\,km), producing the
    sharp peak in pulse delay.  We found no evidence for any kind of
    pulse intensity variations, as from an eclipse, near conjunction.
    The middle panel shows the best-fit residuals obtained using an
    orbital model that does not account for general-relativistic
    effects.  That the residuals deviate significantly from a random,
    Gaussian distribution of zero mean shows that the Shapiro delay
    must be included to properly model the pulse arrival times,
    especially at conjunction.  In addition to the red GBT/GUPPI
    points, the gray points show the previous ``long-term'' data set
    described in the text.  The dramatic improvement in data quality
    is apparent. Finally, the bottom panel shows the post-fit
    residuals for the fully relativistic timing model (including
    Shapiro delay), which have a root-mean-square residual of
    1.1~$\mu$s and reduced $\chi^2$ value of 1.4 with 2165 degrees of
    freedom.}
\end{figure}

\bibliography{1614}

{\bf Acknowledgements} P.B.D.~is a Jansky Fellow of the National Radio Astronomy
  Observatory.  J.W.T.H.~is a Veni Fellow of The Netherlands
  Organisation for Scientific Research (NWO). We thank Jim Lattimer for
  providing the EOS data plotted in Figure~\ref{fig:eos}, and Paulo
  Freire, Feryal {\"O}zel, and Dimitrios Psaltis for useful
  discussions.

{\bf Competing Interests} The authors declare that they have no
 competing financial interests.

{\bf Correspondence} Correspondence and requests for materials should
 be addressed to P.B.D.~(email: pdemores@nrao.edu).

\begin{table}
  \begin{center}
    \include{1614_par}
  \end{center}
  \caption{\label{tab:pars} Physical parameters for PSR~J1614$-$2230:
    Timing model parameters ({\it top}); quantities derived from timing
    model parameter values ({\it middle}); radio spectral and interstellar
    medium properties ({\it bottom}).  Timing values in parentheses represent
    1-$\sigma$ uncertainty in the final digit, as determined by Markov chain
    Monte Carlo error analysis. $\dag$: These quantities vary stochastically
    on $\gtrsim$1-day timescales.  Values presented here are the average for
    our GUPPI data set.  $\ddag$:  Shown in parentheses are separate values
    for the long-term (first) and new (second) data sets described in the
    text. $\S$: Calculated using the NE2001 pulsar distance
    model\cite{ne2001}.}
\end{table}

\end{document}

%% file: 1614_par.tex
\begin{tabular}{lc}
\hline
Ecliptic Longitude, $\lambda$ (deg) \dotfill & 245.78827556(5) \\
Ecliptic Latitude, $\beta$ (deg) \dotfill & -1.256744(2) \\
Proper Motion in $\lambda$ (mas yr$^{-1}$) \dotfill & 9.79(7) \\
Proper Motion in $\beta$ (mas yr$^{-1}$) \dotfill & -30(3) \\
Parallax (mas) \dotfill & 0.5(6) \\
Pulsar Spin Period (ms) \dotfill & 3.1508076534271(6) \\
Period Derivative (s/s) \dotfill & 9.6216(9)$\times$10$^{-21}$ \\
Reference Epoch (MJD) \dotfill & 53600 \\
Dispersion Measure (pc cm$^{-3}$) \dotfill & 34.4865$^\dag$ \\
Orbital Period (days) \dotfill & 8.6866194196(2) \\
Projected Semi-Major Axis (lt-s)  \dotfill & 11.2911975(2) \\
\first~Laplace Parameter, $e\sin\omega$ \dotfill & 1.1(3)$\times$10$^{-7}$ \\
\second~Laplace Parameter, $e\cos\omega$ \dotfill & -1.29(3)$\times$10$^{-6}$ \\
Companion Mass (\msun)    \dotfill & 0.500(6) \\
Sine of Inclination Angle \dotfill & 0.999894(5) \\
Epoch of Ascending Node (MJD) \dotfill & 52331.1701098(3) \\
Span of Timing Data (MJD) \dotfill & 52469$-$55330 \\
Number of TOAs  \dotfill & 2,206 (454 / 1,752)$^\ddag$\\
RMS TOA Residual ($\mu$s) \dotfill & 1.1\\
\hline
Right Ascension (J2000) \dotfill & 16$^{\rm h}$\;14$^{\rm m}$\;36$\fs$5051(5) \\
Declination     (J2000) \dotfill & -22$^\circ$\;30$\arcmin$\;31$\farcs$081(7) \\
Orbital Eccentricity, $e$ \dotfill & 1.30(4)$\times$10$^{-6}$ \\
Inclination Angle (deg) \dotfill & 89.17(2) \\
Pulsar Mass (\msun) \dotfill & 1.97(4) \\
DM-derived Distance (kpc) \dotfill & 1.2$^\S$ \\
Parallax Distance (kpc) \dotfill & $>$0.9 \\
Surface Magnetic Field ($10^8$\,G) \dotfill & 1.8  \\
Characteristic Age (Gyr) \dotfill & 5.2 \\
Spin-down Luminosity ($10^{34}$\,erg\,s$^{-1}$) \dotfill & 1.2 \\
\hline
Average Flux Density at 1.4~GHz (mJy) \dotfill & 1.2$^\dag$ \\
Spectral Index, 1.1--1.9~GHz  \dotfill & -1.9(1) \\
Rotation Measure (rad m$^{-2}$) \dotfill & -28.0(3) \\
\hline
\end{tabular}